\documentclass[aps,prb,showpacs,twocolumn,showpacs]{revtex4-1}

\usepackage{graphicx,graphics,color,epsfig,epstopdf}% Include figure files
\usepackage{bm}
\usepackage{amsmath,amssymb,amsfonts}
\usepackage{dcolumn}% Align table columns on decimal point
\usepackage{times,xspace}
\usepackage{ulem}

\newcommand{\cY}{{\cal Y}}
\newcommand{\vk}{{\bm k}}
\newcommand{\vv}{{\bm v}}
\newcommand{\vA}{{\bm A}}
\newcommand{\vH}{{\bm H}}
\newcommand{\vR}{{\bm R}}
\newcommand{\be}{\begin{equation}}
\newcommand{\ee}{\end{equation}}

\begin{document}

\title{Field-angle-resolved anisotropy in superconducting CeCoIn$_5$ using realistic Fermi surfaces}

\author{Tanmoy Das,$^1$ A. B. Vorontsov,$^2$ I. Vekhter,$^3$,  and Matthias J. Graf$^1$}
\affiliation{$^1$Los Alamos National Laboratory, Los Alamos, New Mexico 87545\\
$^2$Department of Physics, Montana State University, Bozeman, Montana 59717\\
$^3$Department of Physics and Astronomy, Louisiana State University, Baton Rouge, Louisiana 70803}

\date{\today}

\begin{abstract}
We compute the field-angle-resolved specific heat and thermal conductivity using realistic model band structure for the heavy-fermion superconductor CeCoIn$_5$ to identify the gap structure and location of nodes. We use a two-band tight-binding parametrization of the band dispersion as input for the self-consistent calculations in the quasiclassical formulation of the superconductivity. Systematic analysis shows that modest in-plane anisotropy in the density of states and Fermi velocity
in tetragonal crystals significantly affects the fourfold oscillations in thermal quantities, when the magnetic field is rotated in the basal plane.
The Fermi surface anisotropy substantially shifts the location of the lines in the $H$-$T$ plane, where the oscillations change sign compared to quasicylindrical model calculations. In particular, at high fields, the anisotropy and sign reversal are found even for isotropic gaps. Our findings imply that a simultaneous analysis of the specific heat and thermal conductivity, with an emphasis on the low energy sector, is needed to restrict potential pairing scenarios in multiband superconductors. We discuss the impact of our results on recent measurements of the Ce-115 family, namely Ce$T$In$_5$ with $T$=Co,Rh,Ir.
\end{abstract}

%\pacs{74.20.Rp,74.70.Xa,74.25.Uv,74.25.N-}

%Superconductors - vortex phases 74.25.Uv
%Superconductivity - pairing symmetries 74.20.Rp
%Specific heat of superconductors, 74.25.Bt
%Thermal conduction in superconductors, 74.25.fc
%Phase diagrams superconductivity, 74.25.Dw
%Superconducting materials - heavy-fermion materials 74.70.Tx

\maketitle

\section{Introduction}
Many heavy-fermion and other novel superconductors are thought to possess nodes in the gap function on the Fermi surface.
Since the gap shape is directly related to the symmetry of the pairing interaction, knowing the position of nodes can shed light on possible pairing mechanisms. Magnetic field-angle-resolved specific heat and thermal conductivity experiments are able to provide detailed information about the anisotropy of quasiparticle excitations near the Fermi surface, and hence help
identify the nodal directions in the bulk.\cite{IVekhter1999,Miranovic,Anton,YMatsuda2006}  To implement this procedure it is necessary to have high-precision probes that detect small variations under changes of the direction of the applied field.
A series of remarkable experiments proved already the 
viability of this approach.\cite{TPark2003,YMatsuda2006,An2010,Kittaka2012}
However it has proved non-trivial to interpret these experiments in general.
The oscillations in physical quantities, as a function of the field direction, change sign depending on the magnitude of the applied magnetic field and the temperature.\cite{Anton,AntonI,AntonII}
The location of these inversion lines depends sensitively on the topology of the Fermi surface and the material-specific details of the (multi-) band structure.\cite{IVekhter2008,DasFeSe} Obviously, this calls for the development of  theoretical tools that take material-specific properties into account. Furthermore, it suggests that a quantitative and unambiguous identification of the structure of the superconducting (SC) gap requires the incorporation of realistic Fermi surface (FS) properties.

The unconventional heavy-fermion superconductor CeCoIn$_5$ is an ideal candidate for testing field-angle-resolved probes due to the existence of large high-quality crystals and accessible temperature and field ranges. Early field-angle-resolved thermal conductivity and specific heat measurements were controversial on whether CeCoIn$_5$ has a superconducting gap with $d_{x^2-y^2}$ or $d_{xy}$ symmetry. \cite{Izawa2001,Aoki2004}
Recent specific heat measurements observed the predicted inversion of the oscillations at low temperature.\cite{An2010} This seemed to have settled the dispute in favor of $d_{x^2-y^2}$ pairing symmetry.

In this paper we incorporate first-principles electronic structure calculations to obtain the realistic tight-binding parametrization for Ce-115 (Ce$T$In$_5$ with $T$=Co,Rh,Ir) materials that reproduce the Fermi surface (FS) topology and yield the Fermi velocities, and the density of states (DOS) at the Fermi level. We use this FS parametrization  as input for self-consistent calculations of thermal properties in the extended Brandt-Pesch-Tewordt (BPT) approximation of the quasiclassical Eilenberger equation.\cite{AntonI,AntonII} Use of the tight-binding parametrization allows for a numerically efficient computation, while keeping the essential character of the low-energy band structure that reflects on the hybridization between Ce $4f$ and In $5p$ states. Within this framework, we consider candidate $s$- and $d$-wave order parameters, perform a systematic study of the angle-resolved specific heat coefficient, $\gamma=C/T$, and thermal conductivity,  $\kappa$, in a magnetic field rotating in the Ce-In basal plane.
Finally, we construct a field-temperature phase diagram of the fourfold oscillations.

The main results of our calculations, which are applicable to a wide range of systems with tetragonal point group symmetry, are:
(1) For isotropic gap ($s$-wave) we find that moderate FS anisotropies are sufficient to introduce field-angle-dependent oscillations in the specific heat and thermal conductivity in the superconducting state over a significant range of temperatures and at intermediate to high magnetic fields. In addition, the inversion of the oscillation pattern as a function of temperature shows that oscillations are not simply a direct consequence of the anisotropy of the upper critical field.
Therefore not all such oscillations at intermediate fields can be taken as proof of strong anisotropy in the superconducting gap.
This result agrees with our recent numerical study of the iron-based superconductor $T$Fe$_2$Se$_2$.\cite{DasFeSe}
(2) The complex field-angle dependence of the specific heat and thermal conductivity  for systems with anisotropic Fermi surfaces suggests that comparison of both quantities with material-specific theories is required to identify the pairing symmetry and gap structure. This is already important for materials, where the Fermi surface anisotropy is moderate, as is the case for the Ce-115 family.

The rest of the paper is arranged as follows. In Sec.~II, we present our tight-binding representation of the two FSs for CeCoIn$_5$. Detailed analytical and computational formalism of the field-angle resolved specific heat and thermal conductivity calculations is given in Sec.~III.  The results of the temperature and magnetic field dependence of these quantities, and their relative sign reversal in the field-angle oscillation for $s$-and two $d$-wave pairing symmetries are given in Sec.~IV. Some comparison with the available data for CeCoIn$_5$, CeRhIn$_5$, and CeIrIn$_5$ is also included. Finally, we conclude in Sec.~V.

\begin{figure*}[top]
\rotatebox[origin=c]{0}{\includegraphics[width=0.70\textwidth]{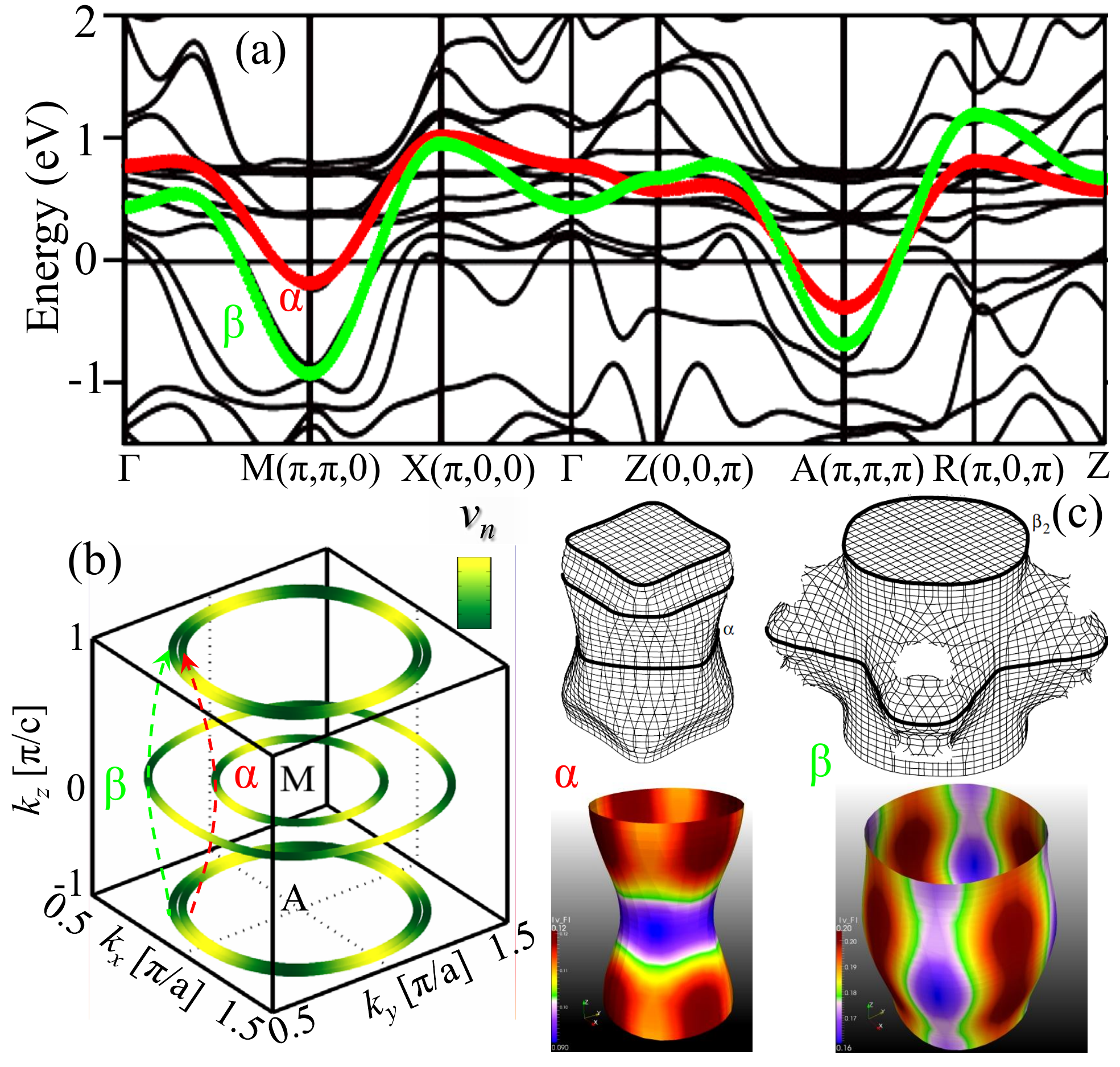}}

\caption{(color online)
Electronic structure of CeCoIn$_5$:
(a) Tight-binding fits of the two most relevant bands $\alpha$ (red) and $\beta$ (green) to the electronic dispersions of CeCoIn$_5$ calculated in the local density approximation by Maehira et al.\cite{Maehira}
(b) $\alpha$ and $\beta$ FSs at three representative $k_z$ values, colored by corresponding Fermi velocity from low (green) to high (yellow).
(c) Three dimensional rendering of the computed FSs for $\alpha$ and $\beta$ bands (bottom panel) compared with the dHvA experiments\cite{Shishido03} (top panel).
The $\alpha$ FS has a narrow waist, while the $\beta$ FS has a belly.
The color map of the calculated FSs gives the anisotropy of the magnitude of the Fermi velocities ranging from low (blue) to high (red).
}
\label{fig1}
\end{figure*}

\section{Electronic structure}
First-principles calculations of CeCoIn$_5$ demonstrate that the bands crossing the Fermi level are dominated by strongly hybridized $4f$ electrons of the Ce atom with weak overlap coming from the $5p$ orbitals of the In atom.\cite{Maehira2003,Maehira} In this work our basic aim is to parameterize the true shape of the FSs only, while the overall dispersion feature at higher energy is irrelevant
for thermodynamic and transport properties. Therefore, we use an effective tight-binding model of the lowest energy of three $4f-$orbitals in a tetragonal lattice. As we are only interested in the eigenvalues and not the eigenvectors of each band, we absorb the orbital symmetry of contributing orbitals into the tight-binding hopping parameters, which makes all bands decoupled from each other. With this motivation we write the tight-binding dispersion including up to third nearest neighbor hopping on the $x-y$ plane and only nearest neighbor hopping along $c-$axis to obtain
\begin{eqnarray}
\xi_{\vk} = -2\sum_i(t_ic_i + t_{2i}c_{2i})-4t_{xy}c_x c_y -E_F.
\end{eqnarray}
Here $c_{\alpha i}=\cos{(\alpha k_i)}$ with $i=x,y,z$. $E_F$ is the Fermi energy. We obtain the values of the tight-binding parameters after fitting to first-principles dispersions  by Ref.~\onlinecite{Maehira} shown in Fig.~1(a):
($t_x$=$t_y$,$t_z$,$t_{2x}$=$t_{2y}$,$t_{xy}$,$E_F$)=(-0.12,-0.05,0,0.09,-0.55), and (-0.17,0.06,0,0.15,-0.47) in eV for the $\alpha$ and $\beta$ band, respectively. Note that $t_{2z}=0$.
The other two bands crossing the Fermi level have small areas and are not further considered in our two-band model description of CeCoIn$_5$.

The $\alpha$ and $\beta$ bands give two concentric electron pockets at the zone corner ($M$-point), see Fig.~1(b). The $k_z$ dispersion of each band is more interesting and needs special attention. Along the $k_z$ direction, both $\alpha$ and $\beta$ FSs are more like corrugated cylinders: the $\alpha$ FS has a narrow waist at $k_z=0$, while the $\beta$ FS has a belly. Note that only nearest neighbor hopping along the $c$ axis is sufficient to obtain the qualitative $k_z$ dispersion of all bands in agreement with the {\it ab-initio} band structure\cite{Shishido} and dHvA experiments\cite{Shishido03} [see Fig.~1(c)]. The opposite sign of the $t_z$ parameter is responsible for the opposite shape of the $\alpha$ and $\beta$ FSs (narrow waist vs.\ belly).

\section{Theory and computational method}

For magnetic field ${\vH}$ applied at angle $\alpha$ with respect to the (100) direction, we compute the field-angle induced superconducting DOS per spin, $N_n(\omega; {\vH})$ (band index $n=1,2$) by solving the Eilenberger equation \cite{Anton,Anton2010,AntonI,AntonII} within the extended BPT quasiclassical approximation. \cite{BPT,WPesch1975,AHoughton1998} The BPT approximation implies a uniform field ${\vH}$ over the unit cell of the Abrikosov vortex lattice (unit-cell averaged Green's function). This produces quantitatively correct results near the upper critical field, and continues to yield semi-quantitatively correct description over the range $0.5 H_{c2}(T) \lesssim H \leq H_{c2}(T)$ for isotropic gap, \cite{BPT,Brandt,Delireu} and to much lower fields for nodal and strongly anisotropic gaps in single-band models.\cite{Vekhter99,TDahm2002,Anton}

Here we summarize the key steps of the calculation, and highlight the main technical differences
between the single- and multi-band systems following Refs.~\onlinecite{AntonI,AntonII,Anton2010}.
The main object of interest, the quasiclassical Green's function, is assumed to be diagonal in the
band space ($n=1,2$), since bands are well separated in the Brillouin Zone, and have the 4$\times$4 Gor'kov-Nambu matrix structure corresponding to singlet pairing in each band,
\be
\hat G =
\left( \begin{array}{cc}
      \hat g_1 & 0 \\
      0 & \hat g_2
   \end{array}
   \right)
   \quad, \quad
\hat g_n =
    \left( \begin{array}{cc}
    g_n & i\sigma_2 f_n \\
    i\sigma_2 \underline{f}_n & -g_n
    \end{array} \right) \,.
\ee
The Green's function in each band satisfies the Eilenberger equation for given Matsubara frequency $i\omega_\nu=i\pi T(2\nu+1)$,
which has a simple commutator form:\cite{Serene1983}
\begin{widetext}
\begin{eqnarray}
   \Big[ (i\omega_\nu + {e\over c} \vv_n(\vk_f) \cdot {\vA}(\vR) )\, \hat{\tau}_3
    - \hat \Delta_n(\vR, \vk_f) - \hat\sigma^{imp}_{n}(i\omega_\nu),
    \,   \hat g_n(\vR, \vk_f; i\omega_\nu)
    \Big] + i\vv_n(\vk_f) \cdot \bm\nabla_\vR \;
    \hat g_n(\vR, \vk_f; i\omega_\nu) = 0 \,,
    \label{eq:eil}
\end{eqnarray}
where the Fermi velocity in band $n$ is denoted by $\vv_n({\vk_f})$, with the wavevector $\vk_f$ on
the respective FS.
Since this is a homogeneous equation, it has to be complemented by the normalization condition of the Green's functions:
\begin{equation}
\hat g_n(\vR, \vk_f; i\omega_\nu)^2 = -\pi^2.
\end{equation}
Furthermore, the off-diagonal Green's functions and self-energies are related by symmetry:\cite{Serene1983}
$\underline{f}_n(\vR, \vk_f; i\omega_\nu) = f_n(\vR, -\vk_f; i\omega_\nu)^* = f_n(\vR, \vk_f; -i\omega_\nu)^*$;
$\underline{\Delta}_n^{imp}(\vR, \vk_f; i\omega_\nu) = \Delta_n^{imp}(\vR, -\vk_f; i\omega_\nu)^* = \Delta_n^{imp}(\vR, \vk_f; -i\omega_\nu)^*$.

The equations for the Green's functions in two bands are coupled indirectly through the
self-energies entering the Eilenberger equation.
The scattering of quasiparticles off impurities with concentration $n_{imp}$ is taken into account via the
self-energy in each band, $\hat\sigma^{imp}_{n}$,
which is evaluated in the $T$-matrix approximation for the two-band system,\cite{Ohashi,Mishra2009}
\be
\hat\sigma^{imp}_{n} \equiv
    \left( \begin{array}{cc}
    D+\Sigma^{imp}_n & i\sigma_2 \Delta^{imp}_n \\
    i\sigma_2 \underline{\Delta}^{imp}_n & D-\Sigma^{imp}_n
    \end{array} \right) = n_{imp} \hat{t}_{nn}
\;,
\qquad
\hat T
= \hat U + \hat U \langle N_{f}(\vk_f) \hat G(\vk_f) \rangle_{FS} \hat T .
\ee
The $\hat T$ matrix and the impurity scattering potential have the following
structure in band space:
\be
\hat T =
\left( \begin{array}{cc}
\hat t_{11}  & \hat t_{12} \\
\hat t_{21}  & \hat t_{22}
 \end{array} \right)
, \qquad
\hat U =
\left( \begin{array}{cc}
u_{11}  & u_{12} \\
u_{21}  & u_{22}
 \end{array} \right) .
\ee
The angular brackets denote the integral over one or the other Fermi surface, as appropriate, e.g.:
\be
\langle \; N_{f}(\vk_f) \, \hat G(\vk_f) \;  \rangle_{FS}
= \mbox{diag}_{n=1,2} \left[ \int_{FS_n}  d^2 k_f \; N_{f,n}(\vk_f) \, \hat g_n(\vk_f) \right]\,,
\ee
and the
corresponding normal-state DOS at the Fermi level is
$N_{f,n}(\vk_f) \sim 1/|\vv_n(\vk_f)|$. Sometimes we will omit the subscript $FS$ for brevity.

For each $T$ and $\vH$ the order parameters
are calculated self-consistently from the coupled gap equations of the two-band model
\begin{eqnarray}
&&\Delta_n(\vR, {\vk_f}) =  T \sum_{\omega_\nu} \sum_{m}
\Big\langle V_{n m}({\vk_f}, {\vk_f}') \;
N_{f,m}({\vk_f}') \,  f_{m}(\vR, {\vk_f}'; i\omega_\nu) \Big\rangle_{FS} .
\label{eq:selfcons}
\end{eqnarray}
\end{widetext}
We use a factorized pairing potential at the Fermi surface as
$V_{n m}({\vk_f}, {\vk_f}') = V_{n m} \; \cY_n(\phi) \, \cY_{m}(\phi')$,
with $\cY_n(\phi)$ the basis function that depends only on the azimuthal angle, see Figs.~\ref{fig2a} and \ref{fig2}.
This means that the order parameters are also factorized,
$\Delta_{1,2}(\vk_f) = \Delta_{1,2} \cY_{1,2}(\phi)$.
We couple the bands, for simplicity, by purely interband pairing $V_{12}=V_{21}=-V$,
and assume same symmetries and angular variations on both bands $\cY_1(\phi)=\cY_2(\phi)$.
This ensures that the order parameters in the two bands are strongly coupled, and the temperature and field dependence of both gaps is similar, while keeping the number of parameters minimal.
While there are indications that in CeCoIn$_5$ there is a small excitation gap that closes at very low fields, of order 0.1\% of the upper critical field,\cite{Seyfarth} it seems likely that this gap is proximity induced on the parts of the Fermi surface with low $f$-electron content that we do not consider here. Since all the experiments measuring the field-angle anisotropy are carried out at fields, which are at sufficiently high $H/H_{c2}$, we  consider only Fermi surface sheets with strong pairing and large gaps.
It is worth mentioning that the interband pairing captures both nodeless $s^\pm$ and nodal $d$ wave pairing scenarios.

Generally, for arbitrary interaction matrix $V_{n m}$,
the coupled gap equations support two solutions for the amplitudes $(\Delta_1, \Delta_2)$.
The physical solution corresponds to the highest transition
temperature $T_{c0}$, that is, the greatest eigenvalue $V_{max}$
of the interaction matrix,
\be
\left( \begin{array}{cc}
V_{11} \langle N_{f,1} \cY_1^2 \rangle & V_{12} \langle N_{f,2} \cY_2^2 \rangle \\
V_{21} \langle N_{f,1}  \cY_1^2 \rangle & V_{22} \langle  N_{f,2} \cY_2^2 \rangle
\end{array} \right)
\left( \begin{array}{c}
e_1 \\ e_2
\end{array} \right)
=
V_{max}
\left( \begin{array}{c}
e_1 \\ e_2
\end{array} \right)
\label{eq:intV}
\,.
\ee
The effective interaction strength $V_{max}$ and the cutoff $\Omega_c$
can be eliminated using standard techniques
in favor of the bare transition temperature, $T_{c0} = 1.13 \Omega_{c} \exp(-1/V_{max})$, \cite{Serene1983}
and the gap amplitudes in different bands are given by the eigenvector of the interaction matrix,
\be
\left( \begin{array}{c}
\Delta_1 \\ \Delta_2
\end{array} \right)
=
\left( \begin{array}{c}
e_1 \\ e_2
\end{array} \right)
\Delta
\:.
\ee
Upon projecting out this vector from Eq.~(\ref{eq:selfcons}),
the system of the self-consistency equations is reduced to a single equation for
the order parameter $\Delta$ of the dominant instability.

Since the only coupling between bands is via the self-consistency equations of the order parameter and the self-energies, the solutions for the propagators in each band can be formally obtained from the transport equation (\ref{eq:eil}) with given $\Delta_n$ and $\sigma_n$ in the same way as for single-band systems.\cite{AntonI,AntonII}
We express the gradient term via the raising and lowering operators $(a^\dag, a)$ for the vortex solutions corresponding to the superposition of different harmonic oscillator functions:\cite{AHoughton1998}
$\vv_n(\vk_f) \cdot \left( \bm\nabla_\vR - i\frac{2e}{c} \vA(\vR) \right) =
    \frac{1}{\sqrt{2} \Lambda} \,\left( v_{n,-}(\vk_f) a^\dagger-v_{n,+}(\vk_f) a \right).
$
The $(x,y)$ projections of the Fermi velocity on the plane perpendicular to the direction of the
field $\hat z$ have to be rescaled by the anisotropy factor $S_f$,
\be
    v_{n,\pm}=v_{n,x}(\vk_f)/\sqrt{S_f}\pm i v_{n,y}(\vk_f)\sqrt{S_f} .
\ee
The relevant parameter that determines the excitations in the SC state at a particular point on the
Fermi surface is the component of the (rescaled) Fermi velocity normal to the applied field,
\be
    v_n^\perp (\vk_f) =
    \sqrt{v_{n,x}(\vk_f)^2/S_f + v_{n,y}(\vk_f)^2 \, S_f } .
\ee
The corresponding energy scale is
\be
\bm{\bar{v}}_{f,n}(\phi,{\vH})\equiv
\frac{{\vv}_n^\perp({\vk_f})}{2\Lambda} ,
\ee
where $\Lambda=(\hbar c/2|e|H)^{1/2}$ is the
magnetic length, which is of order of the intervortex distance, and $\phi$ is the FS angle
with respect to the $k_x$ axis.
The anisotropy parameter $S_f$ is chosen to give the correct form
of the vortex lattice in the linearized Ginzburg-Landau (GL) equations for $\Delta$.
This allows us to consider only the lowest Landau level,\cite{AntonI}
\be
  \Delta(\vR)= \Delta
   \sum_{k_y} C_{k_y}^{(n)} {e^{ik_y\sqrt{S_f} y}
    \over \sqrt[4]{S_f \Lambda^2}} \;
	\Phi_0\left( {x-\Lambda^2 \sqrt{S_f} k_y\over \Lambda \sqrt{S_f}} \right) \,.
\ee

For tetragonal symmetry this parameter depends on the angle that
the applied field makes with the symmetry axis $c$ (in this paper $\theta_H=\pi/2$),
\be
    S_f = \sqrt{ \cos^2 \theta_H +
    {K_{||}\over K_{\perp}} \sin^2 \theta_H}  \,.
\ee
Here $K_{||}$ (along $c$-axis) and $K_{\perp}$ (in-plane) are the coefficients of the gradient terms in the GL expansion
for the gradients along the $c$-axis and in the $ab$-plane respectively. For our two-band system they
depend on the degree of mixing of the bands in a particular superconducting
state $\Delta$. For the state
$(\Delta_1, \Delta_2) = (e_1 \Delta, e_2 \Delta)$
they are determined by the right, $\vec e = (e_1, e_2)^T$, and left, $\vec{e'} = (e_1', e_2')$,
eigenvectors of the interaction matrix in Eq.~(\ref{eq:intV}), corresponding to eigenvalue $V_{max}$ with $\vec{e'} \cdot \vec e = 1$.

\begin{eqnarray}
K_{||} =
\frac{(e'_1 \,,\, e'_2)}{V_{max}}
\left( \begin{array}{cc}
V_{11} & V_{12}  \\
V_{21}  & V_{22}
\end{array} \right)
\left( \begin{array}{c}
e_1\, \langle v_{1c}^2 \, N_{f1} \cY_1^2 \rangle  \\
e_2\, \langle v_{2c}^2 \, N_{f2} \cY_2^2 \rangle
\end{array} \right) ,
\\
K_{\perp} =
\frac{(e'_1 \,,\, e'_2)}{V_{max}}
\left( \begin{array}{cc}
V_{11} & V_{12}\\
V_{21} & V_{22}
\end{array} \right)
\left( \begin{array}{c}
e_1 \, \langle v_{1a}^2 \, N_{f1} \cY_1^2 \rangle \\
e_2 \, \langle v_{2a}^2 \, N_{f2} \cY_2^2 \rangle
\end{array} \right) .
\end{eqnarray}

With these remarks in mind, we can directly use the single-band results for the unit-cell averaged Green's
functions in the single Landau level approximation
[we follow the notation of Eqs.~(46)-(48) in Ref.~\onlinecite{AntonI}]:
\begin{eqnarray}
&g_n(i\omega_{\nu}, {\vk_f}; {\vH}) =
\frac{-i\pi}{
 \sqrt{ 1- i \sqrt{\pi} \frac{1}{\bar{v}_{f,n}^2} W^{\prime}
 \left( \frac{i\tilde\omega_{\nu,n} }{ \bar{v}_{f,n}}  \right) \tilde\Delta_n \underline{\tilde\Delta}_n }
} ,&
\label{eq:gf}
\\
&f_n(i\omega_{\nu}, {\vk_f}; {\vH}) = i g_n
 \frac{\sqrt{\pi} }{\bar{v}_{f,n}} W \left( \frac{i\tilde\omega_{\nu,n} }{ \bar{v}_{f,n}}  \right) \tilde\Delta_n .&
\end{eqnarray}
Here $i\tilde\omega_{\nu,n} = i\omega_{\nu}-\Sigma^{imp}_{n}(i\omega_{\nu}, \vk_f; \vH) $ and
$\tilde\Delta_n=\Delta_n(\vk_f) + \Delta^{imp}_n(i\omega_{\nu}, \vk_f; \vH)$ are the
Matsubara frequency and the order parameter renormalized by the impurity
self-energies in each band $n$, $\hat\sigma^{imp}_{n}$.
$W^{\prime}(z)$ is the first derivative of the complex-valued function $W(z)=\exp{(-z^2)}{\rm erfc}(-iz)$.
One can further cast this in a form similar to that of a uniform superconductor
by introducing the new self-energy  $\Sigma_n$ according to
$i\sqrt{\pi}/{\bar{v}_{f,n}^2} \;  W_n^{\prime}(i\tilde\omega_\nu/\bar{v}_{f,n}) \equiv (i\omega_\nu-\Sigma_n)^{-2} $.
The effective self-energy $\Sigma_n$ now contains effects from both the
impurity scattering and the effects of orbital magnetic field.

In contrast to the Doppler shift approximation, both the real and the imaginary
parts of $\Sigma_n$ contribute to the SC DOS, and their interplay as a function
of energy, $\vH$ and $T$, determine the sign reversal in the fourfold oscillation
of the SC DOS.  These effects have been extensively studied earlier using a
single quasi-cylindrical FS and nodal gap, and a minimal 2D model for two-band
systems, see for example Refs.~\onlinecite{Anton,AntonI,AntonII,Anton2010}.

The transport and thermodynamic coefficients are calculated by using the retarded Green's functions through analytic continuation,  $i\omega_\nu \to \omega + i0$, in the propagators found above.
We begin with the total electronic specific heat from both bands, $C=C_1+C_2$, which is given by the derivative of the net entropy,
$C=T \left(\partial S/\partial T\right)$.  Because the low-temperature approximation given by\cite{AntonI}
\begin{eqnarray}
C_n(\alpha) \approx
2 \int_{-\infty}^\infty
d\omega \frac{\omega^2 \langle N_n(\omega, {\vk_f}; {\vH}) \rangle_{FS}}{4 T^2 {\rm cosh}(\omega/2T)^2} \,,
\label{eq:C}
\end{eqnarray}
remains valid almost up to the normal-state transition region, it can be employed to describe the behavior of the heat capacity over most of the phase diagram.  Detailed numerical calculations show that the high-temperature sign reversal line is robust, but will be shifted to slightly higher temperatures by approximately $0.1 T_{c0}$ for the FS parametrization considered here compared to the calculations using the low-temperature approximation in Eq.~(\ref{eq:C}).

Next, we consider the total electronic thermal conductivity, which is the sum of the contributions from both bands,
$\kappa=\kappa_1+\kappa_2$, with~\cite{Vekhter99,AntonII}
\begin{eqnarray}
\kappa_{n}^{xx}(\alpha) &=&
2 \int_{-\infty}^\infty \!\!
d\omega \frac{ \omega^2 }{2 T^2 {\rm cosh}(\omega/2T)^2} \\
&\qquad& \times{  \langle v_{n}^x({\vk_f})^2 N_n(\omega, {\vk_f}; {\vH})\tau_{n}(\omega,{\vk_f}; {\bm  H}) \rangle_{FS} }
\,.
\nonumber
\label{eq:kappa}
\end{eqnarray}
\begin{widetext}
Here the field-induced SC DOS per spin in each band,
$N_n(\omega,{\vk_f})/N_{f,n}({\vk_f})=-{\rm Im}\ g^R(\omega, {\vk_f}; {\vH})_n/\pi$, the factor $2$ accounts for the spin degeneracy, and the {\it transport} lifetime is due to both impurity and vortex scattering~\cite{Vekhter99,AntonII,Anton2010}
\begin{eqnarray}
  \frac{1}{2\tau_{{n}} (\omega,{\vk_f}; {\bm  H})} =
    - {\rm Im}\, \Sigma^{imp}_n(\omega,{\vk_f};{\vH})
    +
    \sqrt{\pi}{1 \over |{\bar{\vv}}_{f,n}({\vk_f}; {\vH})|}
    \frac{{\rm Im}\, [g^R_n(\omega,{\vk_f};{\vH}) \, W(\tilde{\omega}/|{\bar{\vv}}_{f,n}({\vk_f}; {\vH})|)]}
    {{\rm Im} \, g^R_n(\omega,{\vk_f};{\vH})} |\tilde\Delta_n({\vk_f}; {\vH})|^2
    \,.
    \label{eq:tauH}
\end{eqnarray}
When $T\rightarrow 0$ we recover the standard expressions for the Sommerfeld coefficient,
$\gamma_n\equiv C_n/T \to {2\over 3}\pi^2 \langle N_n(0, {\vk_f}; {\vH}) \rangle_{FS}$,
and for the linear coefficient of the thermal conductivity
$\kappa_n^{xx}/T \to {1\over 3}\pi^2 \langle v_{n}^x({\vk_f})^2 N_n(0, {\vk_f}; {\vH})\tau_{n}(0,{\vk_f}; {\bm  H}) \rangle_{FS}$.
Since the Green's function, given by Eq.~\eqref{eq:gf}, takes the standard BCS form at $H=0$, we also recover the universal thermal conductivity for gaps with nodes on the FS.\cite{Sun1995,Graf1996,Graf1997,Norman1996,Graf1999}
At low fields the approximation breaks down, but for nodal superconductors it provides a good interpolation from low to high fields, and, in the regime
$1 \ll 1/\tau_{imp}\Delta_{n} \ll H/H_{c2} $ reproduces the well-known $\sqrt{H}$ field-dependence of the density of states in $d$-wave superconductors~\cite{Volovik1993,Kuebert98} up to logarithmic corrections.\cite{Vekhter99,AntonI,AntonII}

Since the function $x^2/{\rm cosh}(x/2)^2$ peaks at $x\sim 2.5T$, the anisotropy of the heat capacity at low temperatures is qualitatively determined by the anisotropy in the total SC DOS, $N(\omega=2.5T,{\vk_f}; {\vH})$. Using the expansion of  the error function, we obtain two limiting values for $W^{\prime}(z)$: $W^{\prime}(0)=2i/\sqrt{\pi}$ and $W^{\prime}(z\gg 1)\approx-i/\sqrt{\pi}z^2$. Thus the SC DOS for each band $n$ becomes
\begin{eqnarray}
N_n(\omega; {\vH}) = \langle N_n(\omega,{\vk_f};{\vH})\rangle_{FS}
\approx
\begin{cases}
\left\langle N_{f,n}({\vk_f})\left[1+ 2 \left(\frac{\widetilde\Delta_n({\vk_f}; {\vH})}{|\bar{\vv}_{f,n}({\vk_f}; {\vH})|}\right)^2\right]^{-1/2}\right\rangle_{FS} ,
&\omega\ll \bar{\vv}_{f,n} , \cr\nonumber
\left\langle N_{f,n}({\vk_f})\left[1- \left(\frac{\widetilde\Delta_n({\vk_f}; {\vH})}{\widetilde\omega}\right)^2\right]^{-1/2}\right\rangle_{FS} ,
&\omega\gg \bar{ \vv}_{f,n}  .\cr
\end{cases}\\
\label{Eq:N}
\end{eqnarray}
\end{widetext}
The first line in Eq.~(\ref{Eq:N}) only makes physical sense when the BPT approximation is valid at low energies, i.e., for nodal and strongly anisotropic gaps. In that case at low $T$ (low energy) and  low fields, where $\Delta_n({\vk_f}; {\vH})$ only weakly depends on the direction of the field, the SC DOS depends predominantly on the orientation of
${\bar{ \vv }}_{f,n}({\vk_f};{\vH})$ relative to the minima of $\Delta_n({\vk_f}; {\vH})$. At $\omega=0$ the inversion of the SC DOS as a function of the field for nodal gaps can be obtained in analogy with Refs.~\onlinecite{UdagawaI,AntonI}.

At higher energies, the second line of Eq.~\eqref{Eq:N} has the BCS form apart from the replacement of the bare energies and gaps by their impurity renormalized counterparts.  Therefore the field-angle variation enters via the anisotropy of these self-energies as well as via the field dependence of the gaps, $\Delta_n({\vk_f}; {\vH})$,  for determining the anisotropy of the upper critical field. The latter effect is only relevant in the vicinity of the transition to the normal state, where the result is valid for both nodal and nodeless gaps, including the fully isotropic situation. Crucially, for anisotropic Fermi surfaces the anisotropy in the self-energies and the order parameter is weighted by the normal-state angle-dependent DOS, $N_{f,n}({\vk_f})$, leading to a complex behavior including the switching of the minima and maxima found in this work.
However, in this regime the energy width of the Fermi weighting factor in the integral exceeds the gap amplitude and a full numerical evaluation is required.  Our results are consistent with the general observations based on such an expansion.

For each pairing symmetry, the coupled order parameters are computed self-consistently at each temperature $T$ and for a given value of the  magnetic field ${\vH}$ applied at the angle $\alpha$ to the (100) direction.
We calculate the field-angle oscillations in the $H$-$T$ phase diagram for a mesh of 35 field points between zero and $H_{c2}$, 100 temperature points from zero to $T_{c0}$, and 31 field-angle points $\alpha$ from zero to $90^\circ$ to extract the anisotropic terms in the heat capacity and  thermal conductivity.
For all the calculations,  we consider purely intraband impurity
scattering, $u_{12}=u_{21}=0$, $u_{11}=u_{22}=u_0$,
in the clean limit, $2\Gamma_1 = 1/\tau_{imp,1} = 2 n_{imp}/\pi N_{f1}= 0.01 \times 2\pi T_{c0}$, where
$T_{c0}$ is the bare transition temperature, $N_{f1}$ is the density of states on the first band ($\alpha$-band),
and the scattering phase shift is
chosen to be $\delta= \arctan(\pi N_{f1} u_0) =\pi/2$  (unitarity limit).

\section{Results}

\subsection{Field-induced superconducting DOS anisotropy and the role of Fermi surface topology}

\begin{figure}[top]
\rotatebox[origin=c]{0}{\includegraphics[width=0.99\columnwidth]{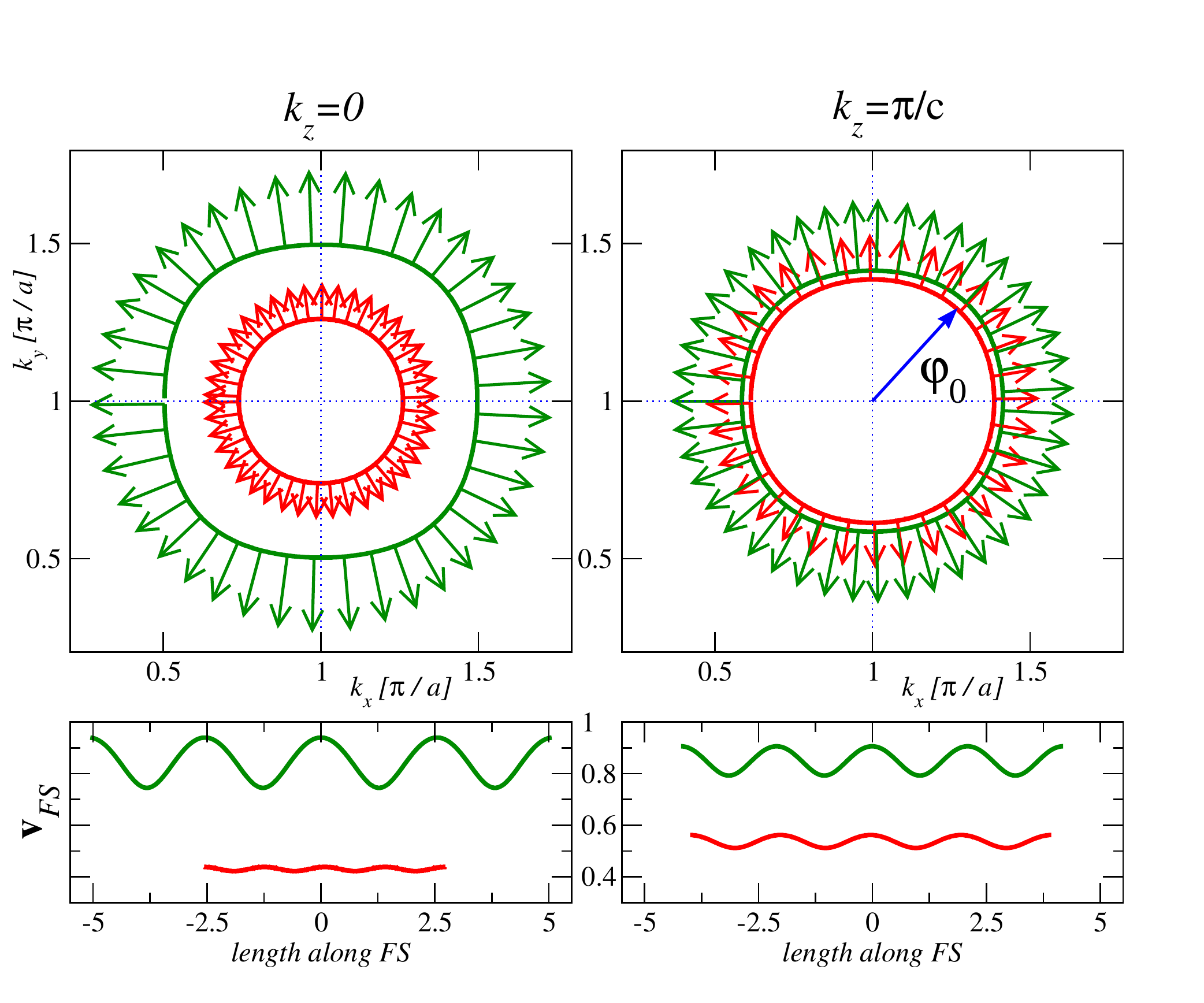}}
\caption{(color online)
Fermi surfaces (FSs) and Fermi velocities at $k_z=0$ (left panel) and $k_z=\pi/c$ (right panel).
The relative magnitude of the Fermi velocities (in arbitrary units) is given by the length of the (red and green) arrows in the top panels.
The bottom panels show the Fermi velocities along the Fermi lines of the corresponding $k_z$ slice.
}
\label{fig2a}
\end{figure}

\begin{figure}[top]
\rotatebox[origin=c]{0}{\includegraphics[width=0.99\columnwidth]{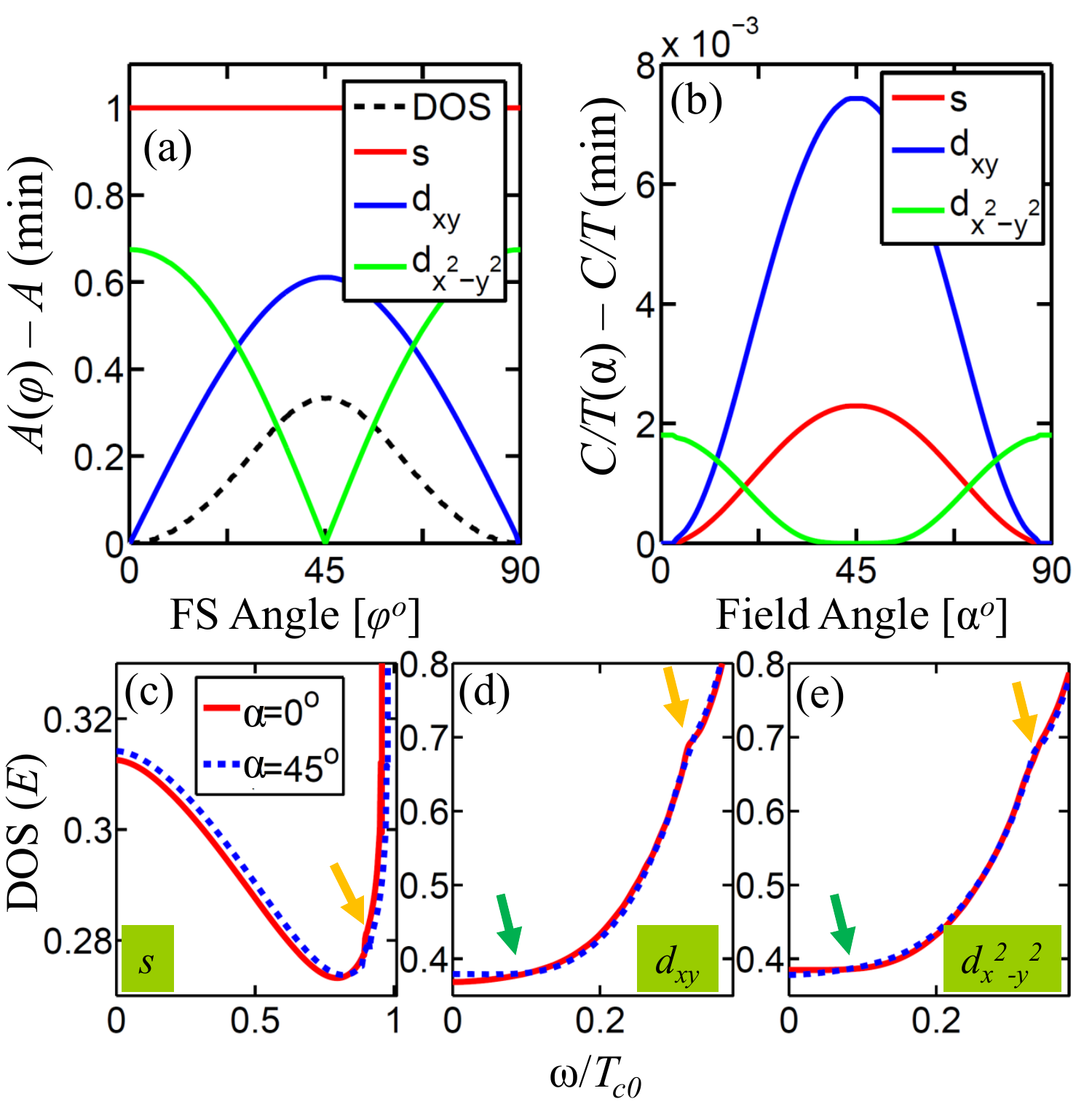}}

\caption{(color online) Fermi surface anisotropy of the normal-state DOS and SC gaps contrasted with the field-angle anisotropy of the Sommerfeld coefficient and the SC DOS.
(a) The calculated FS anisotropy of  the normal-state DOS juxtaposed with
gap functions of three pairing symmetries. All the SC gaps are computed at the FS and all curves are shifted by their corresponding minimum value, except for the $s$-wave gap.
(b) Specific heat coefficient $\gamma(\alpha)=C(\alpha)/T$, normalized to its value at $T_c$,
calculated at $T/T_{c0} = 0.1$ and $H/H_{c2}=0.1$ for $d$-wave gaps and $H/H_{c2}=0.5$ for the $s$-wave gap.
(c)-(e) Field-induced total SC DOS at $T=0$ vs.\  energy at two representative field angles
$\alpha=0^\circ$ and $45^\circ$ for all three pairing symmetries.
Here we used $H/H_{c2}=0.5$ for $s$ wave and 0.1 for both $d$ waves.
Note the low- and high-energy crossings in the SC DOS (arrows) are related to the low- and high-$T$
sign reversals in the oscillations of $\gamma$ and $\kappa$ in Fig.~\ref{fig3}.
}
\label{fig2}
\end{figure}

As discussed above, an important aspect influencing our results is that, for the realistic band structure, the contributions from different segments of the Fermi surfaces to the net density of states are weighted differently according to both the factor $N_{f,n}$ in Eq.~\eqref{Eq:N}, and the segment length of the Fermi surface with a given direction of the Fermi velocity. Fig.~\ref{fig2a} shows the profiles of the Fermi velocity and the corresponding weighting factors. The key point is that, in a tetragonal system,  ${\bm v}_{f,n}$ can have a fourfold anisotropy in the plane that either enhances or competes with the gap anisotropy in determining the contribution to the net DOS in the superconducting state, see Fig.~\ref{fig2}(a). The detailed interplay of the two depends not only on the value of the angle-resolved DOS, but also on the direction of the Fermi velocity.

Indeed, naively one might expect that the relatively large contribution to $N_{f,n}$ from the near-45$^\circ$ direction, combined with the node of the $d_{x^2-y^2}$ at the same angle in Fig.~\ref{fig2}(a) should enhance the field-angle anisotropy for that symmetry of the superconducting state relative to the $d_{xy}$ case when the direction of the greatest $N_{f,n}$ is fully gapped. In fact, the opposite is true, see Fig.~\ref{fig2}(b): the oscillations are enhanced for $d_{xy}$ symmetry.

This is an indication that the flat parts of the Fermi surface with large values
of the Fermi velocity, see Fig.~\ref{fig2a}, contribute more to the total DOS, when the field is at $45^\circ$ and all
four flat parts are 'active'.
When the field is along $0^\circ$ or $90^\circ$, only two flat parts contribute.
In contrast, the four 'active' corners with smaller velocities, and hence slightly larger $N_{f,n}(\vk_f)$
give a smaller contribution simply because their arc length is a smaller fraction of the total Fermi surface length in the respective $k_z$ slice.
It follows that $d_{xy}$-pairing, which has nodes in the flat parts of the FS, exhibits enhanced $C(\alpha=45^\circ)$.
In contrast, the $d_{x^2-y^2}$ profile, gaps those regions of the Fermi surface and thus
anisotropy of $C$ is suppressed. Hence the exact role of the Fermi surface shape and curvature in the field-angle oscillations is highly non-trivial.

At higher temperatures the simple low-$T$ expression in Eq.~(7) is only qualitatively correct, and both detailed calculations \cite{Anton,AntonI,UdagawaI,MiranovicI} and experiments \cite{An2010,FeSeTe_exp,ceirin5}  demonstrated that the anisotropy in the heat capacity is reversed relative to the low-$T$ result.
The lower panel in Fig.~\ref{fig2} shows the field-induced SC DOS as a function of quasiparticle energy below the SC gap for $\alpha=0^\circ$ and $\alpha=45^\circ$ for all three pairing symmetries considered here. We immediately see that the SC DOS at these angles switch and reverse magnitude, which reflects in the sign reversal of the oscillations in specific heat as a function of temperature. Note that, due to the presence of Fermi velocities in $\kappa$ in Eq.~(7), a one-to-one correspondence between SC DOS and $\kappa$ is not straightforward for FS that lack continuous rotational symmetry in the plane.

We show below that for realistic and material-specific anisotropic FS, we still find the sign reversal of the heat capacity oscillations
for $d$-wave pairing, which was previously reported for the rotationally symmetric cases. Hence this sign change is a generic feature of of nodal gaps. However, the key finding in this work is that for moderately anisotropic FSs, measurably large field-angle dependence in the heat capacity and thermal conductivity is obtained at high fields already for {\it isotropic} gaps, which can lead to misinterpretations if analyzed solely in this field range and in terms of simple harmonics of the SC pairing symmetries. We stress that multiband effects add additional complexity to any analysis, due to competing FS anisotropies. For example, we have previously shown that if the FS anisotropy in different bands is opposite (out-of-phase) to each other, then it can lead to additional sign reversals in the field-angle dependence of the thermodynamic quantities for $s$-wave gap, very  similar to what was earlier obtained for nodal gaps only.\cite{DasFeSe}  Furthermore,  bands with different DOS lead to different amplitudes and shapes of the self-consistent value of the SC gaps  (i.e., generally $\Delta_1 \neq \Delta_2$). In this case, the obtained numerical results become less intuitive to interpret and a simple one-to-one mapping between oscillations and nodes is lost. However, at low temperature and low field the generic understanding of the anisotropy as a consequence of the nodal structure alone, remains valid. We give a detailed comparison of the different regimes below.

\subsection{Temperature evolution of field-angle-resolved oscillations}
\begin{figure*}[top]
\rotatebox[origin=c]{0}{\includegraphics[width=0.65\textwidth]{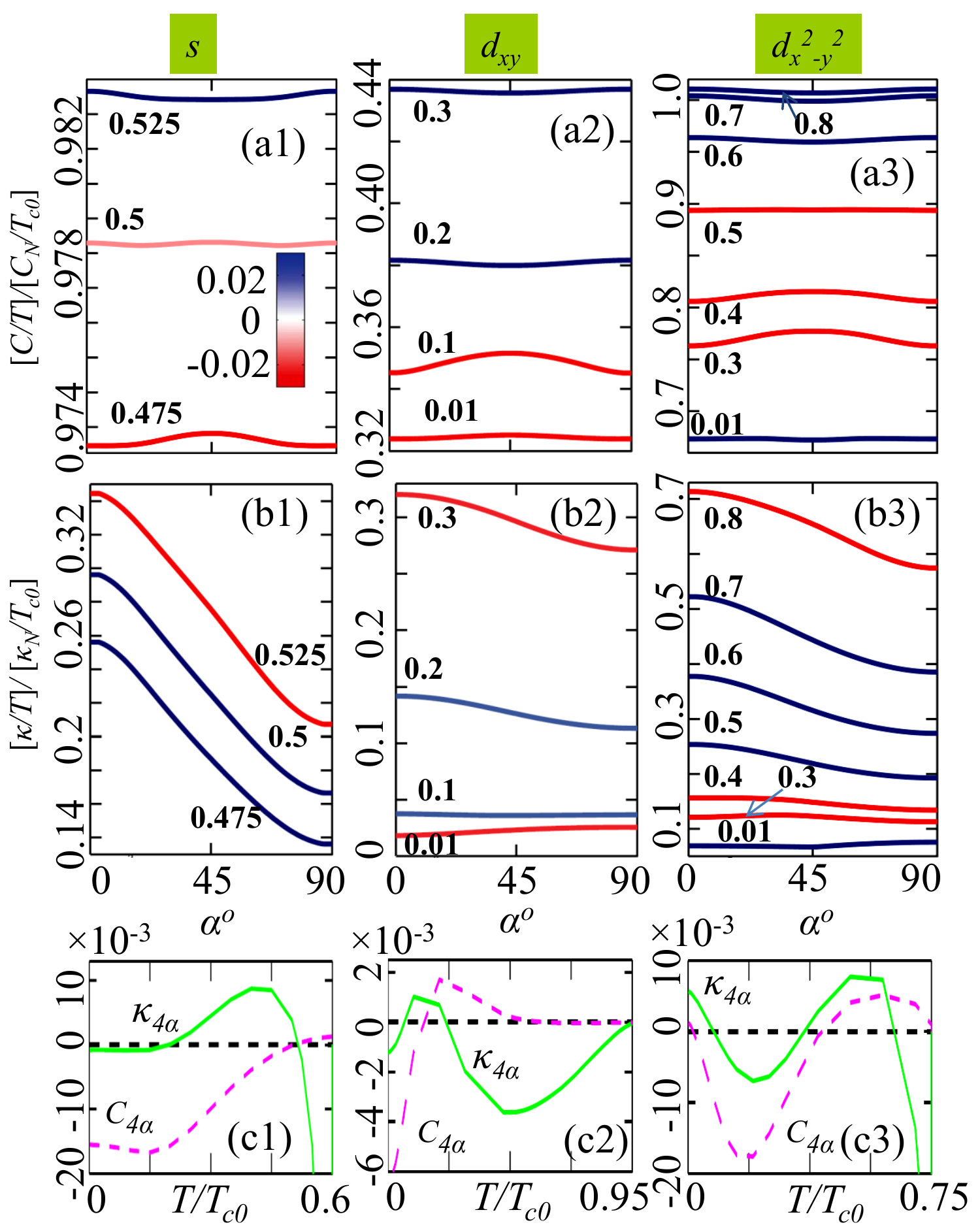}}

\caption{(color online)
Calculated oscillations of the heat capacity and thermal conductivity as a function of the field direction relative to the $x$ axis.
(a1) Sommerfeld coefficient $\gamma=C/T$ normalized to its normal-state value $C_N/T_{c0}$ at fixed field $H/H_{c2}=0.5$ for $s$ wave, plotted from low to high $T$ in units of $T/T_{c0}$
(bottom to top curves).  Each curve is colored by the sign of the fourfold oscillation; a uniform color map is used for values below $-0.025$ and above $0.025$.
(b1) Same as in (a1) but for the normalized thermal conductivity coefficient $\kappa/T$.
(c1) The fourfold  amplitudes of $\gamma$ (dashed line) and $\kappa/T$ (solid line) are plotted as a function of $T$.
For direct comparison the results for nodal $d_{xy}$ ($H/H_{c2}=0.1$) and $d_{x^2-y^2}$ ($H/H_{c2}=0.33$) symmetries are plotted in panels (a2)-(a3), (b2)-(b3), and (c2)-(c3), respectively.
Note that the non-vanishing of $\kappa_{4\alpha}$  as temperature approaches the phase transition line in panels (c1) and (c3) is a consequence of the in-plane anisotropy of $H_{c2}$.
}
 \label{fig3}
\end{figure*}

We present the full angle-dependent profiles of $\gamma(\alpha)=C(\alpha)/T$ and $\kappa(\alpha)/T$ for several temperatures at a representative low field ($H/H_{c2}=0.1$) for the two nodal gaps, and, for comparison, for an isotropic $s$-wave gap at a moderate field ($H/H_{c2}=0.5$) in Fig.~\ref{fig3}. In accordance with our earlier calculation for K$_y$Fe$_{2-x}$Se$_2$ in Ref.~\onlinecite{DasFeSe}, we find that a substantial oscillation in $\gamma$ and $\kappa$ is present for isotropic $s$-wave pairing. The amplitude of the oscillation increases with stronger $k_z$-dispersion. As in simple models,\cite{AntonI} close to the inversion line the oscillations are not a simple sum of the twofold and fourfold harmonics, but have a more complex profile.

For nodal $d_{xy}$ and $d_{x^2-y^2}$ pairings, the behavior of  oscillations of $\gamma(\alpha)$ and $\kappa(\alpha)$ is similar to  results obtained for quasicylindrical FSs, \cite{Anton2010}
however the amplitude of  oscillations and, crucially, the location of  sign reversals  in the $H$-$T$ phase diagram are modified.
Earlier such sign-reversal features were discussed only for highly anisotropic or nodal gap structures.\cite{Anton,AntonI,Anton2010,An2010,FeSeTe_exp,ceirin5} Our material-specific results caution against straightforward interpretation of oscillations at intermediate fields as evidence of nodes, emphasizing the need to probe low energy excitations.

We extract the amplitudes of the fourfold oscillations by defining
\begin{equation}
C_{4\alpha} (T) \equiv \Pi_0^C-\Pi_{45}^C ,
\label{c4}
\end{equation}
where $\Pi_{\alpha}^C=[C(\alpha,T)/T]/[C_N/T_c]$ and
\begin{equation}
\kappa_{4\alpha} (T) \equiv [\Pi_0^\kappa+\Pi_{90}^\kappa]/2-\Pi_{45}^\kappa,
\label{k4}
\end{equation}
where $\Pi_{\alpha}^\kappa=[\kappa^{xx}(\alpha,T)/T]/[\kappa^{xx}_N/T_c]$, and $C_N$ and $\kappa_N$ are the corresponding normal-state values at $T_c$. This definition removes any twofold, sixfold, etc., contribution from $\kappa(\alpha)$ originating from the field parallel or perpendicular to the vortex lines.\cite{kappa}
In fact, it is straightforward to show that for any function $f(\alpha) = \sum_{n=0}^M a_{2n} \cos(2n\alpha)$ the definition
in Eq.~(\ref{k4}) projects out any other harmonic contribution up to $M=5$, resulting in
$\kappa_{4\alpha} = 2 a_{4}$.
We verified numerically that the amplitudes of twelvefold and higher order harmonics are negligible.
On the other hand, the definition in Eq.~(\ref{c4}) is less robust, but very convenient. It gives $C_{4\alpha} = 2 a_{4}$, when $a_2=a_6=a_{10}=0$, which is sufficient when sample misalignment is negligible and when used away from the sign reversal line.

\begin{figure*}[bt]
\rotatebox[origin=c]{0}{\includegraphics[width=0.70\textwidth]{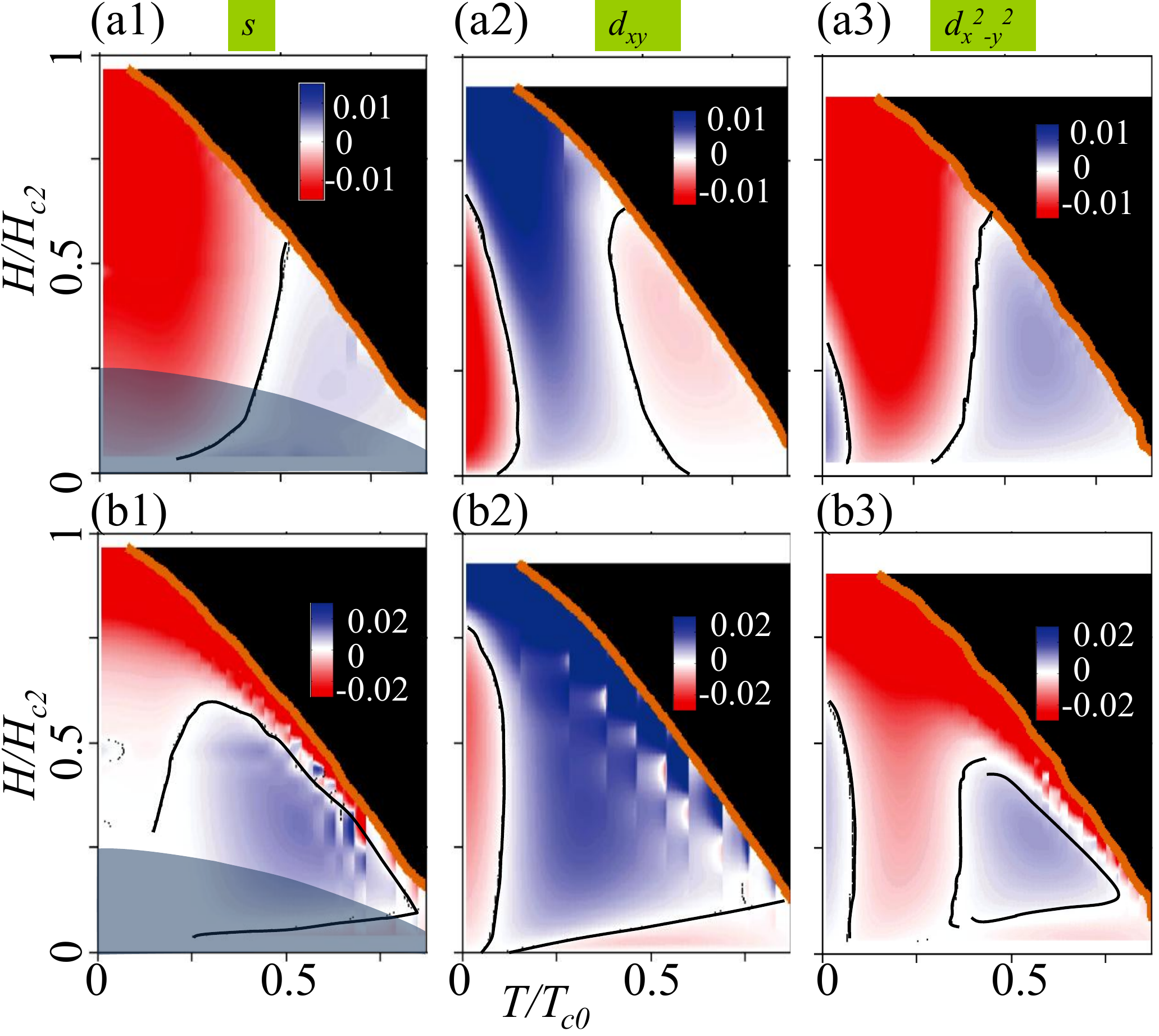}}

\caption{(color online) Contour maps of fourfold amplitude oscillations of normalized specific heat $C_{4\alpha}$ (row $a$) and normalized thermal conductivity $\kappa_{4\alpha}$ (row $b$) in the $H$-$T$ phase diagram. Each column denotes a different gap symmetry studied (isotropic nodeless $s$, nodal $d_{xy}$ and $d_{x^2-y^2}$ gaps). All plots use the same color map scale from minimum (red) to maximum (blue). Note that the fourfold amplitude is given with respect to field ${\vH} \parallel (100)$, i.e., a negative value corresponds to a minimum at $\alpha=0^\circ$.  Here $T_{c}({\vH})$ is defined by the vanishing of both gaps for given symmetry, which determines the line of the (minimum) upper critical field.
Since the BPT approximation for isotropic $s$-wave pairing is not valid at low $H$, we shaded the corresponding area where our approach is not applicable.
} \label{fig4}
\end{figure*}

\subsection{$H$-$T$ phase diagram}

In figure~\ref{fig4} we compile our results of the thermal quantities into a contour map of the amplitude of the fourfold oscillations in the $H$-$T$ phase diagram for $\gamma=C/T$ (top row) and $\kappa/T$ (bottom row) for one nodeless and two nodal gaps. Recall that for quasicylindrical (rotationally-invariant in the basal plane) FSs the specific heat oscillations simply change sign between the $d_{xy}$ and $d_{x^2-y^2}$ symmetries.\cite{AntonI}
While the overall characteristics of the phase diagram remains qualitatively the same for material-specific cases, substantial quantitative changes result from the inclusion of realistic  Fermi surfaces and the directional- and band-dependent contributions to the DOS. Important for the comparison with experiment, we find that the location of the sign-reversal lines for nodal gaps shown in Figs.~\ref{fig4}(a2) and \ref{fig4}(a3) shifted compared to the earlier simple models, due to the interplay of the SC order parameter with the FS anisotropies. As a consequence, the sign of the fourfold oscillations, $C_{4\alpha}$ and $\kappa_{4\alpha}$, may be different over a wider range of temperatures and fields. This is to be contrasted with the results for rotationally symmetric Fermi surfaces, where the two were found to switch sign almost at the same temperatures and fields. We also verified that for $s$-wave pairing the high-$T$ sign reversal is robust and remains at nearly the same location for a single-band superconductor with identical FS.

Note also that at intermediate to high temperatures and fields there is very little in the heat capacity oscillation profile that distinguishes the isotropic gap from that of the $d_{x^2-y^2}$ symmetry, see Fig.~\ref{fig4} panels (a1) vs.\ (a3). However, there is a much more significant difference in transport, Fig.~\ref{fig4} panels (b1) vs.\ (b3), which implies that a simultaneous study of both $C(\alpha)$ and $\kappa(\alpha)$ is highly desirable to gain confidence about the underlying pairing symmetry in any multiband system where the low-temperature, low-field regime is experimentally unreachable. Of course, once the low energy sector at low $T$ and low $H$ is accessed, the differences between different symmetries, and especially between the nodal and isotropic gaps, becomes obvious. Therefore, in general a
rather detailed comparison between measurements and calculations of the $C_{4\alpha}$ and $\kappa_{4\alpha}$ phase diagrams should be employed to draw conclusions about the  pairing symmetries.

\subsection{Comparison with experiments}

\begin{figure*}[th]
\rotatebox[origin=c]{0}{\includegraphics[width=0.80\textwidth]{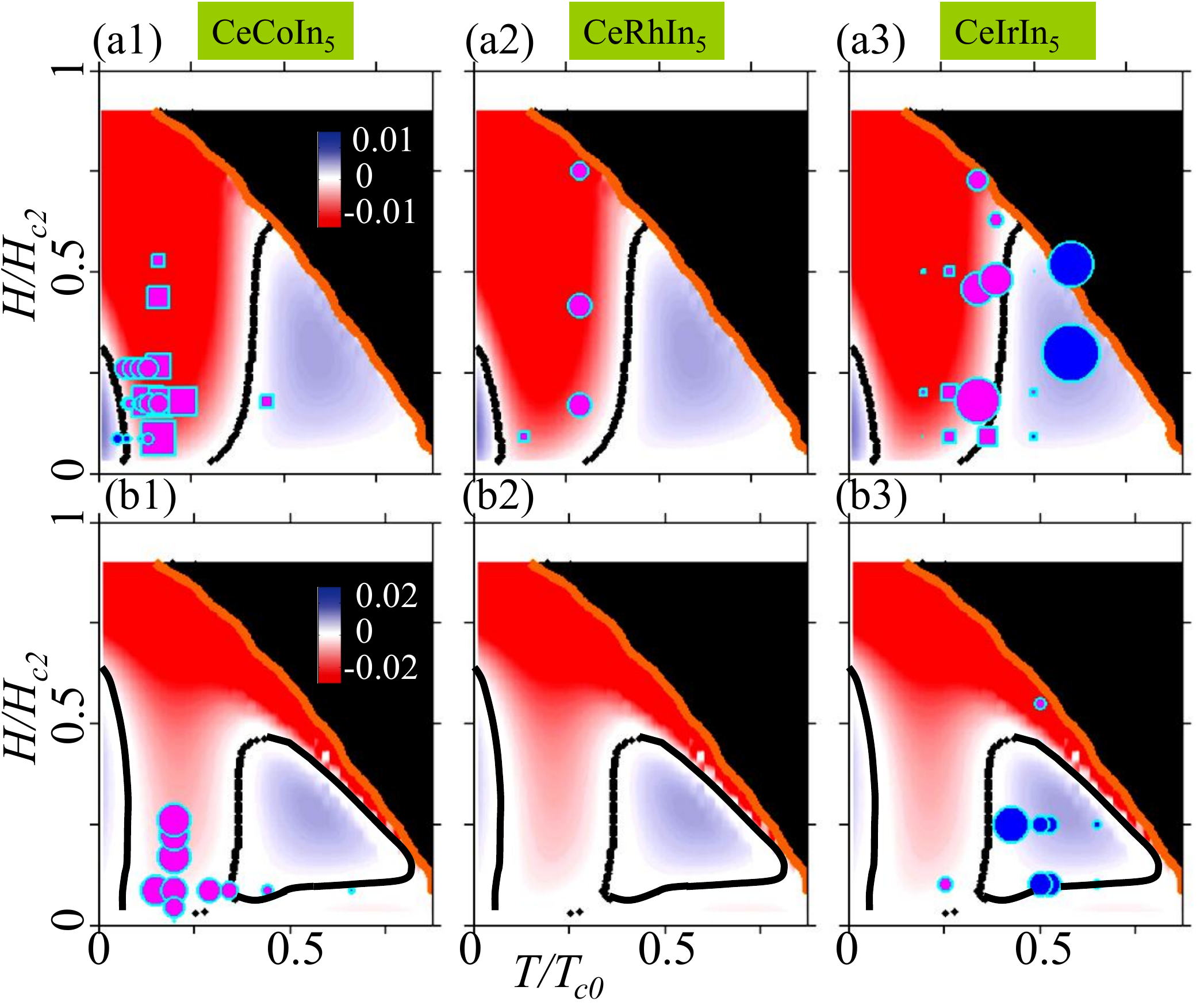}}

\caption{(color online)
Theoretical contour maps of fourfold amplitude oscillations for $d_{x^2-y^2}$ gap for specific heat (upper panel), and thermal conductivity (lower panel) of CeCoIn$_5$, reproduced from Figs.~\ref{fig4}(a3)-b(3). The same theoretical data is repeated in three columns, but compared with three different experimental data for isostructural superconductors within the Ce-115 family.
(a1): Specific heat data of CeCoIn$_5$ by An et al.\cite{An2010} (circles) and Aoki et al.\cite{Aoki2004} (squares).
(b1): Thermal conductivity data of CeCoIn$_5$ by Izawa et al.\cite{Izawa2001}
(a2): Specific heat data for CeRhIn$_5$ by Park et al. \cite{Park2008,Park2009}
(a3): Specific heat data for CeIrIn$_5$ by Lu et al.\cite{ceirin5} (circles) and Kittaka et al.\cite{Kittaka2012} (squares).
(b3): Thermal conductivity data of CeIrIn$_5$ by Kashara et al. \cite{Kasahara2008}
The symbol size gives the corresponding reported amplitude of the oscillation, whereas the filled color depicts its sign. We find reasonable agreement between theory and experiment in both sign and amplitude of oscillations.
} \label{fig6}
\end{figure*}

The superconducting Ce-115 compounds are well suited for the study of field-angle oscillations. Accordingly, there have been a number of experiments investigating the anisotropy of the thermal conductivity and the heat capacity under the rotated field. Here we compare the experimental results with our findings, previously shown in Fig.~\ref{fig4}.
Since the upper critical field is Pauli-limited, and our calculation does not account for the Zeeman splitting, we cannot expect our results to map directly onto the measurements near $H_{c2}$. Nevertheless we believe that a qualitative comparison can be made, especially for systems with strong paramagnetism in the low-field part of the phase diagram, which, when rescaled to the appropriate values of the upper critical field, is essentially identical to that computed in the absence of the Zeeman term.~\cite{Anton2010a}

\par{{\it CeCoIn$_5$:\ }}
The unconventional nature of superconductivity was recognized early on through the discovery of power-law dependence in the temperature behavior of the specific heat and thermal conductivity,\cite{Petrovic2001a,Movshovich2001}
magnetic penetration depth, \cite{Ormeno2002, Chia2003, Ozcan2003, Bauer2006}
and spin-lattice and muon-spin relaxation rates \cite{Kohori2001, Higemoto2002}
consistent with predictions for nodal lines in the gap.
On symmetry grounds the anisotropy of the upper critical field vanishes  near $T_{c0}$, $H_{c2}(0^\circ) = H_{c2}(45^\circ)$. In our calculations, we find that noticeable anisotropy
emerges for $T/T_{c0} \alt 0.7$. In this range $H_{c2}(0^\circ) > H_{c2}(45^\circ)$ for both $s$ and $d_{x^2-y^2}$ pairing symmetries, while the anisotropy is opposite for $d_{xy}$ pairing, i.e., the nodal directions have a lower $H_{c2}$ value. The anisotropy for $s$ and $d_{x^2-y^2}$ gap is in qualitative agreement with the $H_{c2}$ measurements of CeCoIn$_5$ by Settai et al.,\cite{Settai2001} who reported $H_{c2}(0^\circ) > H_{c2}(45^\circ)$ at low temperatures. The measured anisotropy is only a few percent, which would be consistent with the assumption that the band electron $g$-factor, and hence the Pauli limiting field is isotropic in the plane, and the weak anisotropy is due to a residual orbital effect. Remarkably, the opposite relationship,  $H_{c2}(0^\circ) = 11.8\, {\rm T} < H_{c2}(45^\circ)=11.9\, {\rm T}$ was found in Ref.~\onlinecite{Murphy2002}.
So far the experimental discrepancy of the in-plane $H_{c2}$ anisotropy remains an open puzzle.

The original interpretations of the field-angle-resolved thermal conductivity \cite{Izawa2001} and specific heat \cite{Aoki2004} measurements contradicted each other regarding the location of the $d$-wave nodal lines in CeCoIn$_5$. The controversy was finally settled by the observation of the inversion in the specific heat oscillations by An et al.\cite{An2010}
In Fig.~\ref{fig6}(a1) and (b1) we plot both the $C_{4\alpha}$ and $\kappa_{4\alpha}$ experimental data points (symbols). The agreement between theory and experiment is quite convincing for $d_{x^2-y^2}$-wave symmetry and rules out pairing scenarios of either $s$ or $d_{xy}$ gap.

\par{{\it CeRhIn$_5$:\ }}
The  high-pressure, angle-resolved specific heat measurements of CeRhIn$_5$ by Park et al.\cite{Park2008}  showed a clearly delineated fourfold oscillation with $C(0^\circ) < C(45^\circ)$, which was interpreted in favor of $d$-wave symmetry. The measurements were performed down to temperatures as low as 0.3 K ($0.3 T/T_{c}$) and in fields between 0.2 and 0.9 T. At the pressure of 1.47 GPa the superconductivity coexists with antiferromagnetism with superconducting  transition  $T_c = 1.04$ K and in-plane $H_{c2}=1.2$ T at 0.3 K.
The measured in-plane $H_{c2}$ anisotropy was negligible.
As we noted before, in this region of the $H$-$T$ phase diagram both $s$-wave and $d_{x^2-y^2}$-wave gaps are nearly indistinguishable giving rise to fourfold oscillations with the minimum of $C(\alpha)$ occurring at ${\vH} \parallel (100)$. Supporting the $d$-wave interpretation, $T$- and $H$-dependent measurements down to 0.3 K and 0.05 T exhibited power-law behavior consistent with unconventional superconductivity with nodes, i.e., $C/T \sim T$ and $C/T\sim \sqrt{H}$, respectively.
Additional experiments at higher pressure (2.3 GPa), i.e., in the purely superconducting phase, and at $T/T_c=0.14$ and $H/H_{c2} = 0.09$ showed evidence of fourfold oscillations with a negative amplitude $C_{4\alpha}$ of order 4\%.\cite{Park2009}
However, to unequivocally rule out the possibility of  $s$-wave pairing, based on field-angle-resolved measurements alone, experiments would have to be performed at temperatures significantly below $T_c/3$, where the exponential $T$-dependence of the fully gapped excitation spectrum becomes visible.
Power laws were also seen in other pressure measurements of the specific heat, spin-lattice and muon-spin relaxation rates down to $T/T_c \approx 0.15$.\cite{Phillips2002,Mito2001, Higemoto2002}
In chemically doped CeRh$_{1-x}$Ir$_x$In$_5$ a $T^3$ dependence was seen in $1/T_1$ just below $T_c$, which tends toward linear in $T$ at lower temperatures as is typical of dirty $d$-wave superconductors.\cite{Kawasaki2006}
In Fig.~\ref{fig6}(a2) we plot the $C_{4\alpha}$ experimental data points (symbols) on top of the phase diagram for $d_{x^2-y^2}$ gap. The field-angle-dependent experiments taken by themselves are inconclusive, though combined with the reported $T$ and $H$ dependences are strongly suggestive of $d_{x^2-y^2}$-wave superconductivity in CeRhIn$_5$.

\par{{\it CeIrIn$_5$:\ }}
There is an ongoing controversy about the pairing symmetry in this compound, because of its two different superconducting domes, namely one as a function of Rh doping and the other as a function of pressure. In addition, there is disagreement over the interpretation of the thermal conductivity data.
On one side, the field-angle-resolved measurements \cite{Kasahara2008} and power-law dependence in temperature were argued as evidence for $d$-wave gap with vertical line nodes, similar to the sister compound CeCoIn$_5$,\cite{Petrovic2001b,Movshovich2001}
while on the other side thermal conductivity measurements, in particular the temperature and magnetic field dependence of the residual value of $\kappa/T$ along different axes,
were interpreted in favor of a three-dimensional hybrid gap with a horizontal line node. \cite{Shakeripour2007,Shakeripour2009,Shakeripour2010}
The hybrid gap proposal was inspired by similar gap functions studied some time ago for the heavy-fermion superconductor UPt$_3$.\cite{Graf1997,Graf1999,Graf2000}
To further complicate the interpretation, the results by Shakeripour et al.\ were also argued to be  consistent with vertical line nodes.\cite{Vekhter2007}
In addition, power laws were reported for magnetic penetration depth and spin-lattice-relaxation rate.\cite{Vandervelde2009, Kawasaki2006, Kohori2001}
The temperature behavior of the anisotropic penetration depth was interpreted to be consistent with vertical line nodes but not with point nodes and a horizontal line node of the hybrid gap.\cite{Vandervelde2009}
However, the conclusive evidence for the in-plane gap variation comes from
very recent angle-resolved specific heat measurements at ambient and finite pressure.
Lu et al. \cite{ceirin5} (circles) reported fourfold oscillations inside the pressure dome of CeIrIn$_5$ with sign reversal of the oscillations at high temperatures between 0.4 and $0.6 T_{c}$. These data taken together with a low-$T$  anisotropy of $H_{c2}(0^\circ) > H_{c2}(45^\circ)$ and the fact that this compound
belongs to the same family of Ce-115s  was strongly suggestive of two-dimensional $d_{x^2-y^2}$-wave pairing with vertical line nodes. 
Unfortunately, the  temperature in Ref.~\onlinecite{ceirin5} 
was too high to formally exclude isotropic $s$-wave pairing, see the phase diagram in Fig.~\ref{fig4}(a1) versus (a3).
The specific heat data of Ref.~\onlinecite{Kittaka2012}, on the other hand,  were taken down to 80 mK, that is $0.2 T_c$.
Therefore, the specific heat oscillations are supportive of the $d_{x^2-y^2}$ gap scenario. 
Data from both experiments are included in the comparison in Fig.~\ref{fig6}(a3).
In addition field-angle-resolved thermal conductivity data were reported by Kasahara et al.,\cite{Kasahara2008} which are shown in Fig.~\ref{fig6}(b3).
Combined with the specific heat oscillations, they provide strong support for this pairing symmetry.
Hence at present the overwhelming majority of experiments supports the $d_{x^2-y^2}$-wave superconductivity with vertical line nodes in CeIrIn$_5$.

\section{Discussion and Conclusions}

We performed realistic model calculations of the field-angle-resolved specific heat and thermal conductivity using a tight-binding parametrization of the electronic structure within a two-band model of superconductivity, which is relevant for the Ce-115 heavy fermions.
Our systematic analysis of field-angle dependence showed that modest anisotropies in the density of states and the in-plane Fermi velocities of a tetragonal crystal contributes significantly to the fourfold oscillations in the vortex state, when the magnetic field is rotated in the basal plane. As evidence we showed that such oscillations exist at intermediate to high fields even for an isotropic $s$-wave gap. Remarkably, the sign reversal of  fourfold oscillations occurs not only for nodal $d$-wave gaps, but also for an isotropic $s$-wave gap as the temperature is decreased. This is one of the main findings of this work and implies that away from the low temperature and low field region it may be difficult to distinguish different pairing symmetries based on the field-anisotropy of a single probe alone.

Finally, we compared the results of the field-angle-resolved calculations within our model with recent experimental data on different members of the Ce-115 family. The behavior of the self-consistently determined thermal quantities for nodal $d_{x^2-y^2}$-wave gap is consistent with experimental reports for CeCoIn$_5$. The same phase diagram is also consistent with specific heat data for CeRhIn$_5$ and CeIrIn$_5$. Since both CeRhIn$_5$ and CeIrIn$_5$ have similar electronic structure as CeCoIn$_5$ near the Fermi energy, we believe that our Fermi surface parametrization is valid for all three compounds. Consequently, very similar phase diagrams should result for all three materials for which material-specific calculations drastically improved the agreement between theory and experiment. The comparison with experimental data is restricted to low fields, since the superconductivity in this material is Pauli-limited,\cite{Bianchi2003a,Bianchi2003b} and there are indications of a quantum critical point in the vicinity of the upper critical field at zero temperature, $H_{c2}(0)$.\cite{Bianchi2003,Paglione2003} Thus  the regime near the upper critical field at low temperatures is beyond the scope of the current treatment. We find that within our realistic model of the Fermi surface parameters the fourfold anisotropy map is in better agreement with experiments on CeCoIn$_5$, if we assume a weak dispersion along the $k_z$ axis. Note that the relatively small anisotropy of the upper critical field in this material does not have direct connection with the anisotropy of the electron dispersion, as it likely stems from the Pauli limiting of superconductivity. Since the electronic band structure is very similar among the Ce-115s near the Fermi energy, we expect that our findings for CeCoIn$_5$ are also relevant for CeRhIn$_5$ and  CeIrIn$_5$ under pressure. Considering that questions remain about the exact superconducting gap structure and potential spin-fluctuation nesting in the Ce-115s,\cite{Ronning2012} a definitive theoretical account of field-angle-resolved measurements is warranted.

We conclude with a note of caution for interpreting field-angle-resolved oscillations. Our self-consistent two-band model calculations demonstrated that simple observations of oscillations and sign reversals in either $C(\alpha)$ or $\kappa(\alpha)$ are not direct evidence for the presence of nodes or minima in the gap structure. Such conclusions can be drawn either from low-energy measurements, or at higher temperatures and field from a comprehensive simultaneous analysis within the same framework of both $C(\alpha)$ and $\kappa(\alpha)$ measurements. Only a systematic analysis of the fourfold oscillations in the $H$-$T$ phase diagram can constrain the space of possible pairing scenarios for a given material.

\begin{acknowledgments}
We thank R.\ Movshovich, A.~V.\ Balatsky, T.\ Park, F.\ Ronning, and J.~D.\ Thompson for many discussions and encouragements. The work at LANL was funded by the U.S.\ DOE under contract No.\ DE-AC52-06NA25396 through the LDRD program (T.D.) and the Office of Basic Energy Sciences (BES), Division of Materials Sciences and Engineering (M.J.G.).  Work at LSU was supported by NSF Grant No.\ DMR-1105339  (I.V.) and at MSU by NSF Grant No.\ DMR 0954342 (A.B.V.). We are grateful to a NERSC computing allocation by the U.S.\ DOE through BES with contract No.\ DE-AC02-05CH11231.
\end{acknowledgments}

\end{document}